\documentclass[a4paper,10pt]{article}
\usepackage[dvips]{graphicx}
\usepackage{amssymb,amsmath}
\numberwithin{equation}{section}

\begin{document}
\title{\bf{Darboux Transformation and Exact Solutions of the Integrable Heisenberg Ferromagnetic  Equation with Self-Consistent Potentials}}
\author{ Z.S. Yersultanova, M. Zhassybayeva, K. Yesmakhanova,  G. Nugmanova and  \\  R. Myrzakulov\footnote{Email: rmyrzakulov@gmail.com} \\ \textit{Eurasian International Center for Theoretical Physics and  Department of General } \\ \textit{ $\&$  Theoretical Physics, Eurasian National University, Astana 010008, Kazakhstan}}


\date{}
\maketitle
\begin{abstract}
Integrable Heisenberg ferromagnetic equations  are an important subclass of integrable systems. The M-XCIX equation is one of a  generalizations of the Heisenberg ferromagnetic equation  and are integrable.  In this paper, the Darboux transformation  of the M-XCIX equation  is constructed. Using the DT, a 1-soliton solution of the M-XCIX  equation is  presented.

\end{abstract}
\vspace{2cm}
\section{Introduction}

During the past decades, there has been an increasing interest in the study of  nonlinear models, especially,  integrable nonlinear differential equations. Such integrable equations admit in particular soliton or soliton-like solutions. The study of the solitons and related solutions  have become one of active areas of research in physics and mathematics. There are several methods to find soliton and other exact solutions of integrable equations, for instance, Hirota method, inverse scattering transformation, bilinear method, Darboux transformation and so on. Among these methods, the Darboux transformation (DT) is a efficient method to construct the exact solutions of integrable equations. 

 Among of integrable systems, the Heisenberg ferromagnetic   equation (HFE) plays an important role in physics and mathematics \cite{Gut}-\cite{myrzakulov-9535}. In particular, it describes nonlinear dynamics of magnets. Also the HFE can reproduce some integrable classes of curves and surfaces in differential geometry \cite{myrzakulov-715}-\cite{myrzakulov-314}. There are several types integrable HFE. In this paper, we construct the DT for the HFE with self-consistent potentials \cite{Chen1}-\cite{myrzakulov-1397}. Using the DT, we provide soliton solutions of the HFE with self-consistent potentials.
 
The paper is organized as follows.  In section 2, the M-XCIX equation and its Lax representation are introduced.   In section 3, we derived the DT of the M-XCIX equation.   Using these Darboux transformations, one soliton solutions are derived in section 4.  Section 5 is devoted to conclusion.
\section{The M-XCIX equation}
The Myrzakulov-XCIX equation (M-XCIX equation)  reads as \cite{R14}-\cite{R15} (about our notation and definitons, see e.g.  \cite{myrzakulov-391}-\cite{myrzakulov-535})
\begin{eqnarray}
iS_{t}+\frac{1}{2}[S, S_{xx}]+\frac{1}{\omega}[S, W]&=&0,\label{3.21}\\
 iW_{x}+\omega [S, W]&=&0,\label{3.23}
\end{eqnarray} 
where $S=S_{i}\sigma_{i}$, $W=W_{i}\sigma_{i}$, $S^{2}=I, \quad W^{2}=b(t)I$, $b(t)=const(t)$, $I=diag(1,1)$,  $[A,B]=AB-BA$, $\omega$ is a real constant and $\sigma_{i}$ are Pauli matrices. The M-XCIX equation  is   integrable by the IST.  Its   Lax representation can be written in the form
 \begin{eqnarray}
\Phi_{x}&=&U\Phi,\label{3.24}\\
\Phi_{t}&=&V\Phi,\label{3.25} 
\end{eqnarray}  
where the matrix operators $U$ and $V$ have  the form 
 \begin{eqnarray}
U&=&-i\lambda S,\label{2.1}\\
V&=&\lambda^2V_2+\lambda V_{1}+\left(\frac{i}{\lambda+\omega}-\frac{i}{\omega}\right)W.\label{2.2} 
\end{eqnarray} 
Here 
\begin{eqnarray}
V_2&=&-2i S,\quad
V_1=0.5[S,S_x],\label{2.1}\\
S&=&\begin{pmatrix} S_3&S^{+}\\S^{-}& -S_3\end{pmatrix},\quad 
W=\begin{pmatrix} W_3&W^{+}\\W^{-}& -W_3\end{pmatrix},\label{2.2} \\
S^{\pm}&=&S_{1}\pm iS_{2}, \quad W^{\pm}=W_{1}\pm iW_{2}. \label{2.2} 
\end{eqnarray} 
At last, we note that if $W=0$ then the M-XCIX equation becomes the usual HFE 
   \begin{eqnarray}
iS_{t}+\frac{1}{2}[S, S_{xx}]=0.\label{3.23}
\end{eqnarray}
   
 \section{Darboux transformation}
 
 \subsection{One-fold DT}
 In this section, we construct  the DT for the M-XICX equation. To do this, let us  consider the following transformation of solutions of the  equations (2.3)-(2.4)
\begin{eqnarray}
\Phi^{\prime}=L\Phi, \label{3.1}
\end{eqnarray}
where 
\begin{eqnarray}
L=\lambda N-I \label{3.2}
\end{eqnarray}
and 
\begin{eqnarray}
N=\begin{pmatrix} n_{11}&n_{12}\\n_{21}&n_{22}\end{pmatrix}. \label{3.5}
\end{eqnarray}
We require  that $\Phi^{\prime}$ satisfies the same Lax representation as (2.3)-(2.4) so that
\begin{eqnarray}
\Phi'_{x} &=& U'\Phi',\label{3.3}\\
\Phi'_{t} &=& V'\Phi', \label{3.4}
\end{eqnarray}
where $U^{\prime}-V^{\prime}$ depend on $S^{\prime}$ and  $W^{\prime}$ as $U-V$ on $S$ and $W$. The matrix  $L$ obeys the  following equations
\begin{eqnarray}
L_{x}+LU &=& U'L,\label{3.6} \\
L_{t}+LV &=& V'L. \label{3.7}
\end{eqnarray}
 These equations yield  the following equations for $N$ 
   \begin{eqnarray}
   N_{x}&=&iS^{'}-iS, \label{3.10}\\
   N_{t}&=&-S'S'_x-\frac{i}{\omega} W'N+\frac{i}{\omega} NW+SS_x \label{3.15}
\end{eqnarray}
and 
\begin{eqnarray}
   S^{'}&=&NSN^{-1}. \label{3.12}\\
W'&=&(I+\omega N)W(I+\omega N)^{-1}. \label{3.20}
\end{eqnarray}
Also we have  the following useful second form of the DT for $S$:
\begin{eqnarray}
   S^{'}=S-iN_{x}.  \label{3.24}
\end{eqnarray}

\subsubsection{One-fold DT in terms of the $N$ matrix}

We now ready to write the DT for the M-XCIX equation in the explicit form. 
 It can be shown that the matix $N$ has the form
\begin{equation}
N=\begin{pmatrix} n_{11} & n_{12}\\ -n_{12}^{*} & n_{11}^{*}\end{pmatrix},   \label{3.36}
\end{equation}
so that
\begin{equation}
 N^{-1}=\frac{1}{n}\begin{pmatrix} n_{11}^{*} & -n_{12}\\ n_{12}^{*} & n_{11}\end{pmatrix}  \label{3.36}
\end{equation}
and \begin{equation}
I+\omega N=-L_{|\lambda=-\omega}=\begin{pmatrix} \omega n_{11}+1 &\omega n_{12}\\ -\omega n_{12}^{*} & \omega n_{11}^{*}+1\end{pmatrix},\quad 
(I+\omega N)^{-1}=\frac{1}{\square}\begin{pmatrix} \omega n_{11}^{*}+1 & -\omega n_{12}\\ \omega n_{12}^{*} & \omega n_{11}+1\end{pmatrix}. \label{3.37}
\end{equation}
Here
\begin{equation}
n=\det{N}=|n_{11}|^2+|n_{12}|^{2}, \quad \square=det(I+\omega N)=\omega^2|n_{11}|^{2}+\omega(n_{11}+n_{11}^{*})+1+\omega^{2}|n_{12}|^{2}. \label{3.38}
\end{equation}
Finally we have
\begin{eqnarray}
S^{'}&=&\frac{1}{n}\begin{pmatrix}S_{3}(|n_{11}|^{2}-|n_{12}|^{2})+S^{-}n_{11}n_{12}^{*}+S^{+}n^{*}_{11}n_{12} & S^{-}n_{11}^{2}-S^{+}n_{12}^{2}-2S_{3}n_{11}n_{12}\\ S^{+}n_{11}^{*2}-S^{-}n_{12}^{*2}-2S_{3}n_{11}^{*}n_{12}^{*}& S_{3}(|n_{12}|^{2}-|n_{11}|^{2})-S^{-}n_{11}n_{12}^{*}-S^{+}n^{*}_{11}n_{12}\end{pmatrix},\label{3.39}
\\
    W'&=&\frac{1}{\square}\begin{pmatrix}1+A_{11} & A_{12}\\ -A_{21}& 1-A_{22}\end{pmatrix}, \label{3.41}
\end{eqnarray}
where 
\begin{eqnarray}
A_{11}&=&(\omega^2|n_{11}|^{2}+\omega(n_{11}+n_{11}^{*})-|n_{12}|^{2})W_3+(\omega n^{*}_{11}+1)n_{12}W^{+}+(\omega n_{11}+1)n^{*}_{12}W^{-},\label{3.42}\\
A_{12}&=&{-2\omega n_{11}n_{12}W_3-2n_{12}W_3+\omega^2n_{11}^2W^{-}+2\omega n_{11}W^{-}+W^{-}-n_{12}^2W^+},\label{3.43}\\
A_{21}&=&-2\omega n^*_{11}n^*_{12}W_3-2n^*_{12}W_3+\omega^2(n^*_{11})^2W^{+}+2\omega n^{*}_{11}W^{+}+W^{+}-(n_{12}^{*})^2W^-,\label{3.44}\\
A_{22}&=&(\omega^2|n_{11}|^{2}+\omega(n_{11}+n_{11}^{*})-|n_{12}|^{2})W_3+(\omega n^{*}_{11}+1)n_{12}W^{+}+(\omega n_{11}+1)n^{*}_{12}W^{-}. \label{3.45}
\end{eqnarray}
At last, we give the another form of the solutions of $S$ as:
\begin{eqnarray}
S^{'}=S-i\begin{pmatrix} n_{11x} & n_{12x}\\ -n_{12x}^{*} & n_{11x}^{*}\end{pmatrix}, \label{3.46}
\end{eqnarray}
 so that
 \begin{eqnarray}
S^{+\prime}&=&S^{+}+in_{12x}^{*}, \label{3.46}\\
S^{-\prime}&=&S^{-}-in_{12x}, \label{3.46}\\
S^{\prime}_{3}&=&S_{3}-in_{11x}. \label{3.46}
\end{eqnarray}
 
 \subsubsection{One-fold DT in terms of eigenfunctions}
 
 Let the column  $(\psi_{1}, \psi_{2})^{T}$ is the solution of Eqs.(2.3)-(2.4)  with $\lambda$. Then the  new column $(-\psi_{2}^{*}, \psi_{1}^{*})^{T}$ is the solution of Eqs.(2.3)-(2.4) as $\lambda^{*}$.  We now consider the following matrix solution
 \begin{eqnarray}
H=\begin{pmatrix} \psi_{1}(\lambda_{1};t,x,y)&\psi_{1}(\lambda_{2};t,x,y)\\\psi_{2}(\lambda_{1};t,x,y)&\psi_{2}(\lambda_{2};t,x,y)\end{pmatrix}. \label{3.48}
\end{eqnarray}
It satisfies the system:
\begin{eqnarray}
H_{x} &=& -iSH\Lambda, \label{3.50}\\
H_{t} &=& -2iSH\Lambda^2+SS_xH\Lambda-\frac{i}{\omega}WH+WH\Sigma, \label{3.51}
\end{eqnarray}
where 
\begin{eqnarray}
\Lambda=\begin{pmatrix} \lambda_{1}&0\\0&\lambda_{2}\end{pmatrix},  \quad \Sigma=\begin{pmatrix} \frac{i}{\lambda_{1}+\omega}&0\\0&\frac{i}{\lambda_{2}+\omega}\end{pmatrix}, \label{3.52}
\end{eqnarray}
$det$ $H\neq0$ and  $\lambda_{k}$  are complex constants. We now assume that  the matrix $N$ can be written as:
\begin{eqnarray}
N=H\Lambda^{-1} H^{-1}. \label{3.47}
\end{eqnarray}
  From these equations follow that $N$ obeys the equations
  \begin{eqnarray}
N_{x} &=& iNSN^{-1}-iS, \label{3.53}\\
N_{t} &=& SS_x-NSS_xN^{-1}-\frac{i}{\omega}(WN-NW)+WH\Sigma\Lambda^{-1}H^{-1}-NWH\Sigma H^{-1}, \label{3.55}
\end{eqnarray}
which are equivalent to Eqs.(3.8)-(3.9) as we expected.  
  In order to satisfy the constraints of $S$ and  $W$, the $S$ and matrix solutions of the system (2.3)-(2.4) obey the condition
\begin{eqnarray}
\Phi^{\dagger}=\Phi^{-1}, \quad S^{\dagger}=S, \label{3.62}
\end{eqnarray} 
which follow from the equations
\begin{eqnarray}
\Phi^{\dagger}_{x}=i\lambda \Phi^{\dagger}S^{\dagger}, \quad (\Phi^{-1})_{x}=i\lambda \Phi^{-1}S^{-1}. \label{3.63}
\end{eqnarray}
Here $\dagger$ denote an Hermitian conjugate. After some calculations we came to the formulas 
\begin{eqnarray}
\lambda_{2}=\lambda^{*}_{1}, \quad
 H=\begin{pmatrix} \psi_{1}(\lambda_{1};t,x,y)&-\psi^{*}_{2}(\lambda_{1};t,x,y)\\ \psi_{2}(\lambda_{1};t,x,y)&\psi^{*}_{1}(\lambda_{1};t,x,y)\\ \end{pmatrix}, \label{3.64}
\end{eqnarray}
\begin{eqnarray}
H^{-1}=\frac{1}{\Delta}\begin{pmatrix} \psi^{*}_{1}(\lambda_{1};t,x,y)&\psi^{*}_{2}(\lambda_{1};t,x,y)\\ -\psi_{2}(\lambda_{1};t,x,y)&\psi_{1}(\lambda_{1};t,x,y)\\ \end{pmatrix}, \label{3.65}
\end{eqnarray}
where 
\begin{eqnarray}
\Delta &=&|\psi_{1}|^2+|\psi_{2}|^2. \label{3.66}
\end{eqnarray}
So finally for the matrix $N$ we get the following expression
\begin{eqnarray}
N&=&\frac{1}{\Delta}\begin{pmatrix} \lambda_{1}^{-1}|\psi_{1}|^2+\lambda^{-1}_{2}|\psi_{2}|^2 & (\lambda_{1}^{-1}-\lambda_{2}^{-1})\psi_{1}\psi_{2}^{*}\\ (\lambda_{1}^{-1}-\lambda_{2}^{-1})\psi_{1}^{*}\psi_{2} & \lambda_{1}^{-1}|\psi_{2}|^2+\lambda_{2}^{-1}|\psi_{1}|^2)\end{pmatrix}.  \label{3.67}
\end{eqnarray}
Hence we can write the DT in terms of the eigenfunctions of the Lax representations (2.3)-(2.4) as
\begin{eqnarray}
S^{+\prime}&=&S^{+}+i\left(\frac{(\lambda_{1}^{-1}-\lambda_{2}^{-1})\psi_{1}^{*}\psi_{2}}{\Delta}\right)_{x},
\label{3.68}\\
S^{-\prime}&=&S^{-}-i\left(\frac{(\lambda_{1}^{-1}-\lambda_{2}^{-1})\psi_{1}\psi_{2}^{*}}{\Delta}\right)_{x},
\label{3.68}\\
S_{3}^{\prime}&=&S_{3}-i\left(\frac{\lambda_{1}^{-1}|\psi_{1}|^2+\lambda^{-1}_{2}|\psi_{2}|^2}{\Delta}\right)_{x}
\label{3.68}
\end{eqnarray}
and similarly for $W$.

Lastly let us  unify our notations rewriting  the 1-fold DT as:
\begin{equation}
\Phi^{[1]}=L_1\Phi \label{3.76}
\end{equation}
where
\begin{equation}
L_{1}=\lambda l^{1}_{1}+l_{1}^{0}=\lambda l^{1}_{1}-I\label{3.77}
\end{equation}
and  $l_{1}^{0}=-I$.
Then the 1-fold DT takes the form
\begin{eqnarray}
    S^{[1]}&=& l^{1}_{1}S(l^{1}_{1})^{-1},  \label{3.78}\\
    W^{[1]}&=&L_{1}|_{\lambda=-\omega}WL_{1}^{-1}|_{\lambda=-\omega} \label{3.80}
\end{eqnarray}
or
\begin{eqnarray}
    S^{[1]}&=& L_{1\lambda}S(L_{1\lambda})^{-1},  \label{3.78}\\
    W^{[1]}&=&L_{1}|_{\lambda=-\omega}WL_{1}^{-1}|_{\lambda=-\omega}. \label{3.80}
\end{eqnarray}
\subsection{Two-fold DT}
In this subsection we want give some main formulas of the 2-fold DT. We start from the transformation
\begin{equation}
\Phi^{[2]}=L_2\Phi^{[1]}=(\lambda N_2-I)\Phi^{[1]}=(\lambda N_2-I)(\lambda N_1-I)\Phi=(l_{2}^{0}+\lambda l^{1}_2+\lambda^2 l^{2}_2)\Phi \label{3.81}
\end{equation}
where
\begin{equation}
L_2=l_{2}^{0}+\lambda l^{1}_2+\lambda^2 l^{2}_2. \label{3.82}
\end{equation}
Here
\begin{equation}
l^{1}_2=-(N_1+N_2), \quad l^{2}_2=N_2N_1,  \quad l_2^0=I \label{3.83}
\end{equation}
We have
\begin{eqnarray}
L_{2x}+L_{2}U &=& U^{[2]}L_{2},\label{3.84} \\
L_{2t}+L_{2}V &=& V^{[2]}L_{2}. \label{3.85}
\end{eqnarray}
Then Eq.(3.52) gives the coefficients of $\lambda^{i}$ as:
\begin{eqnarray}
      	\lambda^0&:&  l^{0}_{2x}=0\label{3.87}\\
	 \lambda^1&:& l^{1}_{2x}=il^{0}_{2}S-iS^{[2]}l^{0}_{2}, \label{3.88}\\
 \lambda^2&:& l^{2}_{2x}=il^{1}_{2}S-iS^{[2]}l^{1}_{2}. \label{3.89}\\
 \lambda^3&:& 0=il^{2}_{2}S-iS^{[2]}l^{2}_{2}. \label{3.90}
\end{eqnarray}
Hence in particular we get
\begin{eqnarray}
   S^{[2]}=l^{2}_{2}S(l^{2}_{2})^{-1}. \label{3.91}
\end{eqnarray}
 We need in the following formulas:
 \begin{eqnarray}
  \frac{\lambda}{\lambda+\omega}=1- \frac{\omega}{\lambda+\omega}, \quad \frac{\lambda^{2}}{\lambda+\omega}=\lambda-\omega+\frac{\omega^{2}}{\lambda+\omega}.\label{3.92}
\end{eqnarray}
 Then  coefficients of $\lambda^{i}$ of  two sides of the  equation (3.53) give us the following equations
\begin{eqnarray} 
\lambda^{0}&:& l^{0}_{2t}-\frac{i}{\omega}l^{0}_{2}W+il^{1}_{2}W-i\omega l^{2}_{2}W=-\frac{i}{\omega}W^{[2]}l^{0}_{2}+iW^{[2]}l^{1}_{2}-i\omega W^{[2]}l^{2}_{2}, \label{3.14}\\
\lambda^{1}&:& l^{1}_{2t}+l^{0}_{2}SS_{x}-\frac{i}{\omega}l^{1}_{2}W+il^{2}_{2}W=S^{[2]}S^{[2]}_xl^{0}_{2}-\frac{i}{\omega} W^{[2]}l^{1}_{2}+iW^{[2]}l^{2}_{2}, \label{3.15}\\
\lambda^{2}&:&l^{2}_{2t}-2il^{0}_{2}S+l^{1}_{2}SS_{x}-\frac{i}{\omega}l^{2}_{2}W=-2iS^{[2]}l^{0}_{2}+S^{[2]}S^{[2]}_xl^{1}_{2}-\frac{i}{\omega}W^{[2]}l^{2}_{2},\label{3.16}\\
\lambda^{3}&:&-2il^{1}_{2}S+l^{2}_{2}SS_{x}=-2iS^{[2]}l^{1}_{2}+S^{[2]}S^{[2]}_{x}l^{2}_{2},\label{3.16}\\
\lambda^{4}&:&-2il^{2}_{2}S=-2iS^{[2]}l^{2}_{2},\label{3.16}\\
(\lambda+\omega)^{-1}&:&il^{0}_{2}W-i\omega l^{1}_{2}W+i\omega^{2}l^{2}_{2}W=iW^{[2]}l^{0}_{2}-i\omega W^{[2]}l^{0}_{2}-i\omega W^{[2]}l^{1}_{2}+i\omega^{2}W^{[2]}l^{2}_{2}.\label{3.17} 
\end{eqnarray}
Hence we obtain the following 2-fold DT:
\begin{eqnarray} 
S^{[2]}&=&l^{2}_{2}S(l^{2}_{2})^{-1},\label{3.16}\\
W^{[2]}&=&(l^{0}_{2}-\omega l^{1}_{2}+\omega^{2}l^{2}_{2})W(l^{0}_{2}-\omega l^{1}_{2}+\omega^{2} l^{2}_{2})^{-1}\label{3.17} 
\end{eqnarray}
and
\begin{eqnarray} 
 l^{0}_{2t}=0. \label{3.14}\end{eqnarray}
 Note these equations  we can rewrite as
 \begin{eqnarray} 
S^{[2]}&=&L_{2\lambda\lambda}S(L_{2\lambda\lambda})^{-1},\label{3.16}\\
W^{[2]}&=&L_{2}|_{\lambda=-\omega}W(L_{2}|_{\lambda=-\omega})^{-1}\label{3.17} 
\end{eqnarray}
and 
\begin{eqnarray} 
 L_{2t}|_{\lambda=0}=0, \label{3.14}\end{eqnarray}
 respectively.
\subsection{n-fold DT}

Let us now  we  construct the n-fold DT. It has the form
\begin{equation}
\Phi^{[n]}=L_n\Phi^{[n-1]}=(\lambda N_n-I)\Phi^{[n-1]}=(\lambda N_n-I)\ldots(\lambda N_2-I)(\lambda N_1-I)\Phi \label{3.113}
\end{equation}
so that
\begin{equation}
\Phi^{[n]}=[\lambda^n l_{n}^n+\lambda^{n-1} l_{n}^{n-1}+ ... +\lambda l^{1}_{n}+l_{n}^{0}]\Phi, \label{3.114}
\end{equation}
where $l_{n}^{0}=(-1)^{n}I$.
The $n$-fold DT of the M-XCIX equation can be  written as:
\begin{eqnarray}
L_{nx}&=&U^{[n]}L_n-L_{n}U, \label{3.115}\\
L_{nt}&=&V^{[n]}L_{n}-L_{n}V. \label{3.116}
\end{eqnarray}
Hence, in particular, we obtain 
\begin{eqnarray} 
S^{[n]}&=&\frac{\partial^{n} L_{n}}{\partial \lambda^{n}}S\left(\frac{\partial^{n} L_{n}}{\partial \lambda^{n}}\right)^{-1},\label{3.16}\\
W^{[n]}&=&L_{n}|_{\lambda=-\omega}W(L_{n}|_{\lambda=-\omega})^{-1}\label{3.17} 
\end{eqnarray}
and
\begin{eqnarray} 
 L_{nt}|_{\lambda=0}=0. \label{3.14}\end{eqnarray}


\section{Soliton solutions}

 Now we consider a seed solution 
 \begin{eqnarray}
    S=\sigma_{3},  \quad W=b\sigma_{3}, \label{4.17}
\end{eqnarray}where $b=const(t)$. Then we get
\begin{eqnarray}
    S^{[1]}&=& \frac{1}{n}\begin{pmatrix}|n_{11}|^{2}-|n_{12}|^{2} & -2n_{11}n_{12}\\ -2n_{11}^{*}n_{12}^{*}& |n_{12}|^{2}-|n_{11}|^{2})\end{pmatrix}, \label{4.18}\\
W^{[1]}&=&\frac{1}{\square}\begin{pmatrix}{b(\omega^2|n_{11}|^{2}+\omega(n_{11}+n_{11}^{*})+1-|n_{12}|^{2}} & -2\omega n_{11}n_{12}-2n_{12})\\ 
    -2\omega n^{*}_{11}n^{*}_{12}-2n^{*}_{12}& -b(\omega^2|n_{11}|^{2}+\omega(n_{11}+n_{11}^{*})+1-|n_{12}|^{2})\end{pmatrix}. \label{4.20}
\end{eqnarray}
Now we  are ready to write the solutions of the M-XCIX equation  in terms of the elements of $N$. We get
\begin{eqnarray}
    S^{+[1]}&=&-\frac{2n_{11}^{*}n_{12}^{*}}{n} ,  \label{4.21}\\
   S^{-[1]}&=&-\frac{2n_{11}n_{12}}{n},\label{4.22}\\
    S^{[1]}_{3}&=&\frac{|n_{11}|^{2}-|n_{12}|^{2}}{n}, \label{4.23}\\
          W^{+[1]}&=&\frac{-2bn^*_{12}(\omega n^*_{11}+1)}{ \square}, \label{4.25} \\
   W^{-[1]}&=&\frac{-2bn_{12}(\omega n_{11}+1)}{\square},\label{4.26}\\
W^{[1]}_{3}&=&\frac{(\omega^2|n_{11}|^{2}+\omega(n_{11}+n_{11}^{*})+1-|n_{12}|^{2})b}{\square}. \label{4.27}
    \end{eqnarray}
 In our case the eigenfunctions are given by
\begin{eqnarray}
\psi_{1}&=&e^{-i\lambda x+i[-2\lambda^2+b(\frac{1}{\lambda+\omega}-\frac{1}{\omega})]t+i\delta_{1}}=e^{\theta_{1}+i\chi_{1}},\label{4.32}\\
\psi_{2}&=&e^{i\lambda x+i[2\lambda^2-b(\frac{1}{\lambda+\omega}-\frac{1}{\omega})]t+i\delta_{2}}=e^{\theta_{2}+i\chi_{2}},\label{4.33}\end{eqnarray}
where
 \begin{eqnarray}
 \theta_1&=&\beta x-2\alpha\beta t+\frac{\beta b}{(\alpha+\omega)^2+\beta^2}t-\sigma_1,\label{4.49}\\
 \theta_2&=&-\beta x-2\alpha\beta t-\frac{\beta b}{(\alpha+\omega)^2+\beta^2}t-\sigma_2,\label{4.49}\\
 \chi_1&=&-i\alpha x-2i\alpha^2 t-i\beta^2 t+\frac{ib(\alpha+\omega)}{(\alpha+\omega)^2+\beta^2}t-\frac{i}{\omega}bt+i\tau_1,\label{4.49}\\
 \chi_2&=&i\alpha x+2i\alpha^2 t-i\beta^2 t-\frac{ib(\alpha+\omega)}{(\alpha+\omega)^2+\beta^2}t+\frac{i}{\omega}bt+i\tau_2.
 \end{eqnarray}
where $\lambda=\alpha+i\beta$ and $\delta_i=\tau_i+i\sigma_i$. Here $\alpha,\beta,b,\tau,\sigma, \delta_0$ are real constants. In this case the elements of $N$ have the form  
\begin{eqnarray}
  n_{11}&=&\frac{1}{\alpha^2+\beta^2}(\alpha-i\beta\tanh2\theta_1),\label{4.49}\\
  n_{12}&=&\frac{-i\beta e^{2i\chi_1-i\delta_0}}{(\alpha^2+\beta^2)\cosh2\theta_1},
\end{eqnarray}and
 \begin{eqnarray}
\square=\omega^{2}+2\alpha\omega+\alpha^{2}+\beta^{2}=(\omega+\alpha)^{2}+\beta^{2}, \label{4.39}
\end{eqnarray}
Finally the 1-soliton solution of the M-XCIX equation has the form
\begin{eqnarray}
S^{[1]}_{3}&=&\tanh^22\theta_1+\frac{\alpha^2-\beta^2}{\alpha^2+\beta^2}\frac{1}{\cosh^22\theta_1}=\frac{1}{\cosh^22\theta_1}\left(\sinh^22\theta_1+\frac{\alpha^2-\beta^2}{\alpha^2+\beta^2}\right), \label{4.49}\\
S^{+[1]}&=&\frac{2\beta}{\alpha^2+\beta^2}e^{-i\chi_1+i\chi_2}\left(\frac{\beta\sinh2\theta_1}{\cosh^22\theta_1}-\frac{i\alpha}{\cosh 2\theta_1}\right), \label{4.49}\\
 S^{-[1]}&=&\frac{2\beta}{\alpha^2+\beta^2}e^{i\chi_1-i\chi_2}\left(\frac{\beta\sinh 2\theta_1}{\cosh^22\theta_1}+\frac{i\alpha}{\cosh2\theta_1}\right),
\end{eqnarray}
 
 Similarly, we can find solutions for $W$. They have the form
\begin{eqnarray}
W^{[1]}_{3}&=&\frac{b}{M}\left[\frac{\omega^2}{\cosh^22\theta_1}\left(\sinh^22\theta_1+\frac{\alpha^2-\beta^2}{\alpha^2+\beta^2}\right)+2\alpha\omega+\frac{1}{\alpha^2+\beta^2}\right], \label{4.49}\\
 W^{+[1]}&=&\frac{2\beta\omega e^{-2i\chi_1+i\delta_0}b}{(\alpha^2+\beta^2)M\cosh2\theta_1}\left[\beta\omega\tanh 2\theta_1-i(\alpha\omega+\alpha^2+\beta^2)\right], \label{4.49}\\
  W^{-[1]}&=&\frac{2\beta\omega e^{2i\chi_1-i\delta_0}b}{(\alpha^2+\beta^2)M\cosh2\theta_1}\left[\beta\omega\tanh 2\theta_1+i(\alpha\omega+\alpha^2+\beta^2)\right],
\end{eqnarray}
where
\begin{equation}
M=(\omega+\alpha)^2+\beta^2.
\end{equation}
Also we note that take place the following formula (which may be will useful to understand the nature of the matrix function $W$):
\begin{eqnarray}
W^{[1]}=\frac{\omega^2b}{M}S^{[1]}+F^{[1]}, \label{4.49}
\end{eqnarray}
where
 \begin{eqnarray}
F^{[1]}_{3}&=&\frac{b}{M}\left(2\alpha\omega+\frac{1}{\alpha^2+\beta^2}\right), \label{4.49}\\
F^{+[1]}&=&-\frac{2i\beta\omega b e^{-2i\chi_1-i\delta_0}}{M\cosh2\theta_1}, \label{4.49}\\
F^{-[1]}&=&\frac{2i\beta\omega b e^{2i\chi_1+i\delta_0}}{M\cosh2\theta_1}.
\end{eqnarray}

 \section{Conclusion}

 The Heisenberg ferromagnetic   equations  are fascinating  nonlinear dynamical models. In particular, integrable  Heisenberg ferromagnetic   equations  have much relevance in applied ferromagnetism and nanomagnetism. Integrable  Heisenberg ferromagnetic   equations have a close connection to the nonlinear Schrodinger models as well as to the differential geometry of curves and surfaces (see e.g. \cite{R15} and references therein). So it is considered to be one of the fundamental integrable equations admitting  an $N$-soliton solution. 
 
In soliton theory, the construction of the analytical and simple form of the
solutions for the integrable partial and ordinary differential equations is one of central topics in
recent years. In this context, up to now the Darboux transformation method, originated from the work of Darboux in 1882 for the Sturm-Liouville equation,
is a powerful method to construct exact solutions for integrable linear and nonlinear differential equations.   Various approaches have been
proposed to find a Darboux transformation for a given integrable system, for example, the gauge transformation method, the loop group transformation, the operator factorization
method. Indeed,
through iterations, the Darboux transformation is often leaded to compact representations in terms of special determinants
like  as Grammian or Wronskian   for the exact $N$-soliton solutions of the given equation. Such $N$-soliton solutions are appealing
both form practical application viewpoint and form the theoretical viewpoint  as well.

Since integrable systems have remarkable mathematical properties and numerous physical
applications, their generalizations or extensions have attracted attention of many
researchers. One possible direction is extensions with self-consistent sources. This sort of extensions
may also be physically interested. The most famous example might be the nonlinear Schrodinger - Maxwell - Bloch  equation, which now is one of the most important equations in theory
of pulse propagation along the optical fiber.
Another interesting fundamental soliton equation is the Heisenberg ferromagnetic   equation with self-consistent sources \cite{R14}-\cite{R15}.

 In this paper, we have derived the DT of the M-XCIX equation including one-fold, two-fold and n-fold DT.  This equation describe a nonlinear dynamics of the (1+1)-dimensional ferromagnets with self-consistent potentials and is integrable. As an example, the 1-soliton solution of the M-XCIX equation have been  constructed explicitly by using the DT from some trivial seed solution. Of course, in a recursive manner, we can construct the $N$-soliton solution as well as the other type nonlinear solutions like: breathers,  positons and so on. Moreover,  this constructed  Darboux transformation, in particular, allows us to calculate higher order rogue wave solutions of the M-XCIX equation and other  Heisenberg ferromagnetic   equations with self-consistent sources. We will study these important solutions of the M-XCIX equation as well as  its integrable generalizations  in future.

 \end{document}